\documentclass[longauth,letter,traditabstract]{aa} 
\usepackage{graphicx}
\usepackage{txfonts}
\begin{document}
   \title{First results of {\em Herschel} / PACS observations of Neptune                           \thanks{{\em Herschel} is an ESA space observatory with science instruments provided
by European-led Principal Investigator consortia and with important participation from NASA.}
}

\author{E. Lellouch\inst{1}
\and P.~Hartogh\inst{2}
\and H.~Feuchtgruber\inst{3}
\and B.~Vandenbussche\inst{4}
\and T.~de Graauw\inst{5}
\and R.~Moreno\inst{1}
\and C.~Jarchow\inst{2}
\and T.~Cavali\'e\inst{2}
\and G.~Orton\inst{6}
\and M.~Banaszkiewicz\inst{7}
\and M.I.~Blecka\inst{7}
\and D.~Bockel\'ee-Morvan\inst{1}
\and J.~Crovisier\inst{1}
\and T.~Encrenaz\inst{1}
\and T.~Fulton\inst{8}
\and M.~K\"{u}ppers\inst{9}
\and L.M.~Lara\inst{10}
\and D.C.~Lis\inst{11}
\and A.S.~Medvedev\inst{2}
\and M.~Rengel\inst{2}
\and H.~Sagawa\inst{2}
\and B.~Swinyard\inst{12}
\and S.~Szutowicz\inst{7}
\and F.~Bensch\inst{13}
\and E.~Bergin\inst{14}
\and F.~Billebaud\inst{15}
\and N.~Biver\inst{1}
\and G.A.~Blake\inst{6}
\and J.A.D.L.~Blommaert\inst{4}
\and J.~Cernicharo\inst{16}
\and R. Courtin\inst{1}
\and G.R. Davis\inst{17}
\and L.~Decin\inst{4}
\and P.~Encrenaz\inst{18}
\and A.~Gonzalez\inst{2}
\and E.~Jehin\inst{19}
\and M.~Kidger\inst{20}
\and D.~Naylor\inst{21}
\and G. Portyankina\inst{22}
\and R.~Schieder\inst{23}
\and S.~Sidher\inst{12}
\and N.~Thomas\inst{22}
\and M.~de Val--Borro\inst{2}
\and E.~Verdugo\inst{20}
\and C.~Waelkens\inst{4}
\and H.~Walker\inst{12}
\and H. Aarts \inst{5}
\and C. Comito\inst{24}
\and J.H. Kawamura\inst{6}
\and A. Maestrini\inst{18}
\and T.~Peacocke\inst{25}
\and R. Teipen\inst{23}
\and T. Tils\inst{23}
\and K. Wildeman\inst{5}
}

\institute{LESIA, Observatoire de Paris, 5 place Jules Janssen, F-92195 Meudon, France\\ 
\email{emmanuel.lellouch@obspm.fr}
\and Max-Planck-Institut f\"{u}r Sonnensystemforschung, Katlenburg-Lindau, Germany           
\and Max-Planck-Institut f\"ur extraterrestrische Physik, 
Giessenbachstra\ss e, 85748 Garching, Germany           
\and Instituut voor Sterrenkunde, Katholieke Universiteit Leuven, Belgium                
\and SRON, Groningen, the Netherlands                                                    
\and Jet Propulsion Laboratory, California Institute of Technology, Pasadena, United States                                            
\and Space Research Centre, Polish Academy of Science, Warszawa, Poland                  
\and Blue Sky Spectroscopy Inc., Lethbridge, Alberta, Canada                             
\and European Space Astronomy Center, Madrid, Spain                                        
\and Instituto de Astrof\'isica de Andaluc\'ia (CSIC), Granada, Spain                   
\and California Institute of Technology, Pasadena, United States                        
\and Rutherford Appleton Laboratory, Oxfordshire, United Kingdom                        
\and Deutsches Zentrum f\"{u}r Luft- und Raumfahrt (DLR), Bonn, Germany                 
\and University of Michigan, Ann Arbor, United States                                   
\and Universit\'e de Bordeaux, Observatoire Aquitain des Sciences de l'Univers,
CNRS, UMR 5804, Laboratoire d'Astrophysique de Bordeaux, France                                                   
\and Laboratorio de Astrof\'isica Molecular, CAB. INTA-CSIC, Spain                               
\and Joint Astronomy Center, Hilo, United States                                        
\and LERMA, Observatoire de Paris, and Univ. Pierre et Marie Curie, Paris, France       
\and F.R.S.-FNRS, Institut d'Astrophysique et de G\'eophysique, Li\`ege, Belgium                     
\and Herschel Science Centre, European Space Astronomy Centre, Madrid, Spain                                     
\and University of Lethbridge, Canada                                                   
\and University of Bern, Switzerland                                                    
\and University of Cologne, Germany                                                     
\and Max-Planck-Institut für Radioastronomie, Bonn, Germany                             
\and Experimental Physics Dept., National University of Ireland                         
Maynooth, Co. Kildare. Ireland
}



   \date{Draft \today; received ???; accepted ???}


\abstract{We report on the initial analysis of a {\em Herschel} / PACS full range spectrum of Neptune,
covering the 51-220 $\mu$m range with a mean resolving power of $\sim$3000, and complemented by a dedicated observation
of CH$_4$ at 120 $\mu$m. Numerous spectral features due to HD (R(0) and R(1)), H$_2$O, CH$_4$, and CO are present,
but so far no new species have been found. Our results indicate that (i) Neptune's mean thermal profile  is
warmer by $\sim$3 K than inferred from the {\em Voyager} radio-occultation; (ii) the D/H mixing ratio is (4.5$\pm$1)$\times$10$^{-5}$,
confirming the enrichment of Neptune in deuterium over the protosolar value ($\sim$2.1$\times$10$^{-5}$); (iii) the CH$_4$ mixing ratio in the mid stratosphere
is (1.5$\pm$0.2)$\times$10$^{-3}$, and CH$_4$ appears to decrease in the lower stratosphere
at a rate consistent with local saturation, in agreement with the scenario of CH$_4$ stratospheric injection 
from Neptune's warm south polar region; (iv) the H$_2$O stratospheric 
column is (2.1$\pm$0.5)$\times$10$^{14}$ cm$^{-2}$ but its vertical distribution is still to be determined, so the
H$_2$O external flux remains uncertain by over an order of magnitude; and (v) the CO stratospheric
abundance is about twice the tropospheric value, confirming the dual origin
of CO suspected from ground-based millimeter/submillimeter observations.

 }


\authorrunning{Lellouch et al.}

\titlerunning{PACS observations of Neptune}

\maketitle
%

\section{Introduction}

Neptune's thermal emission has been initally explored from the ground in the 8-13 $\mu$m window and in the millimeter range and by the {\em Voyager} spacecraft in 1989, but detailed views of its spectrum had to await sensitive instrumentation onboard ISO (see review in B\'ezard et al. 1999a), {\em
Spitzer}
(Meadows et al. 2008) and recently AKARI (Fletcher et al. 2010). Altogether, these observations have revealed a surprisingly rich
composition of Neptune's stratosphere, including numerous hydrocarbons (CH$_4$, C$_2$H$_2$, C$_2$H$_6$, CH$_3$, C$_2$H$_4$, 
CH$_3$C$_2$H, C$_4$H$_2$), oxygen-bearing species (CO, CO$_2$, and H$_2$O), HCN, as well as deuterium species CH$_3$D and HD.
Favorable factors for observing minor species in Neptune's atmosphere are (i) its relatively warm stratosphere ($\sim$140~K at 1 mbar)
that enhances IR emission; and
(ii) Neptune's large internal heat source 
that results in rapid convection updrafting minor
disequilibrium species, notably CO, up to observable levels.  Neptune's submillimeter spectrum longwards of 50 $\mu$m
has been observed by ISO/LWS (Burgdorf et al. 2003), but the signal--to--noise ratio in the data was not high enough to reveal spectral features.
In this paper, we report the first results from observations of Neptune at 51-220 $\mu$m (195--45 cm$^{-1}$) with the PACS
instrument onboard {\em Herschel} (Pilbratt et al. 2010), 
performed in the framework of the KP-GT ``Water and Related Chemistry in the Solar System", also known as
``{\em Herschel} Solar System observations" (Hartogh et al. 2009).

\begin{figure*}[ht]
\centering
\includegraphics[height=17.cm,angle=-90]{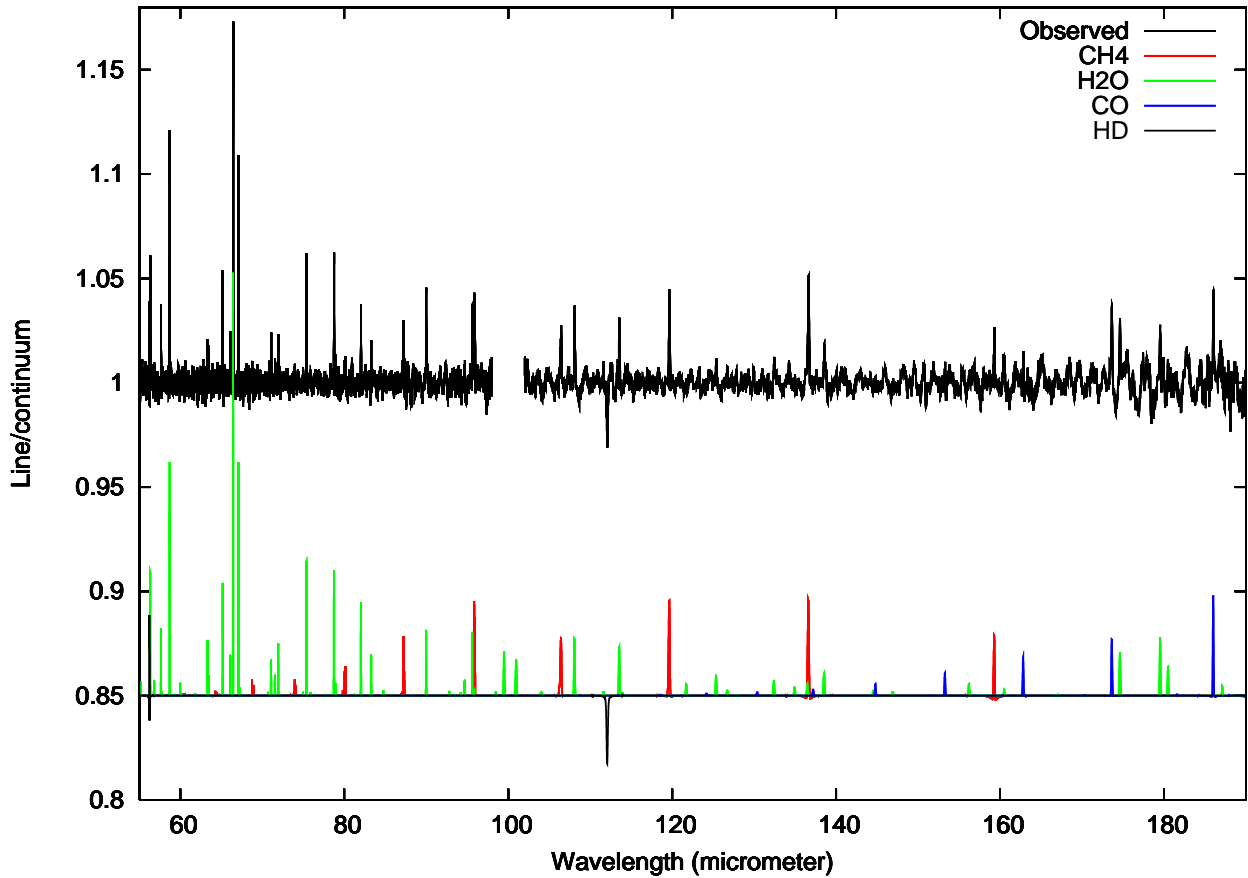}
\caption{Composite PACS spectrum of Neptune, expressed in line/local continuum ratios. For spectral ranges covered more than once (Table~\ref{obssum}), 
the observation with the highest resolution has been selected. The region beyond 190 $\mu$m is not shown, owing to severe mixing
of spectrometer orders. The bottom curves are synthetic spectra at appropriate spectral resolution that show the contributions of CH$_4$, H$_2$O, CO, and HD lines.}
\label{fig:overview}
\end{figure*}


\section{{\em Herschel} / PACS observations}

\begin{table}
\begin{minipage}[t]{\columnwidth}
\caption{Summary of observations} 
\label{obssum}      
\centering                          
\renewcommand{\footnoterule}{} 
\begin{center}
\begin{tabular}{lccc}
\noalign{\smallskip} \hline \noalign{\smallskip}
Obs. ID &  Start Time          &  T$_{obs}$ & Range\footnote{grating order and filter: k = 2A, l = 3A, m = 2B, n = 1 red}\\
           &      [UTC]             &   [min.]     & [$\mu$m] \\
\noalign{\smallskip} \hline \noalign{\smallskip}
1342186536 &  30-Oct-2009 00:58:36   &  116    &  51-72$^k$, 102-145$^n$  \\
1342186537 &  30-Oct-2009 03:01:48   &  133    &  51-62$^l$, 150-186$^n$ \\
1342186538 &  30-Oct-2009 05:22:32   &  203    &  60-73$^l$, 180-220$^n$ \\
1342186539 &  30-Oct-2009 08:53:20   &  151    &  68-85$^m$, 120-171$^n$ \\
1342186540 &  30-Oct-2009 11:31:41   &  236    &  82-102$^m$, 165-220$^n$ \\
1342186571 & 31-Oct-2009 14:35:00& 82 & 118.4-120.9$^n$\\
\hline
\end{tabular}
\end{center}
\end{minipage}
\end{table}

All observations (Table 1) were carried out in chopped-nodded PACS range spectroscopy modes (\cite{poglitsch10}) at high spectral sampling density. 
The entire spectral range of PACS has been measured at full instrumental resolution $\lambda/\delta\lambda$ ranging from 950 to 5500 depending 
on wavelength and grating order (\cite{pacs10}). A summary of the observations is given in Table~\ref{obssum}. 
Since blue and red spectrometer data are acquired in parallel, several spectral ranges have been observed in overlap. 
Given the instrumental spatial pixel size of 9.4"$\times$9.4",  Neptune (2.297" as seen from {\em Herschel})
can be considered as a point 
source, and the analyzed spectra therefore originate only in the central spatial pixel of the integral field spectrometer.

Starting from Level 0 products, the processing of all observations was carried by standard PACS pipeline 
modules (\cite{poglitsch10}) up to Level 1. Individual spectral pixels were then scaled onto their common mean in 
order to improve the removal of signal outliers caused by cosmic ray hits. After application of an iterative $\sigma$-clipping, 
adapted to the instrumental resolution, the remaining data were rebinned onto an oversampled wavelength grid to ensure 
conservation of spectral resolution. The absolute flux calibration of the instrument and improvements on the relative 
spectral response function are still in progress. Therefore the resulting spectrum was then divided by its continuum, 
to be robust against forthcoming calibration updates. The composite spectrum is shown in Fig.\ref{fig:overview}. It shows emission
signatures due to CH$_4$, H$_2$O, CO, as well as the R(0) and R(1) lines of HD at 112 and 56 $\mu$m, seen respectively in absorption and emission.
At this stage of the data reduction, features below $\sim$0.5-1 \% contrast must be treated with caution. No new species are detected at this level.

A dedicated line--scan high S/N observation of the CH$_4$ 119.6 $\mu$m rotational line was also acquired in order to get a high precision measure
of the CH$_4$ stratospheric abundance.

\section{Analysis and discussion}
\begin{figure}[ht]
\centering
\includegraphics[width=6.4cm,angle=-90]{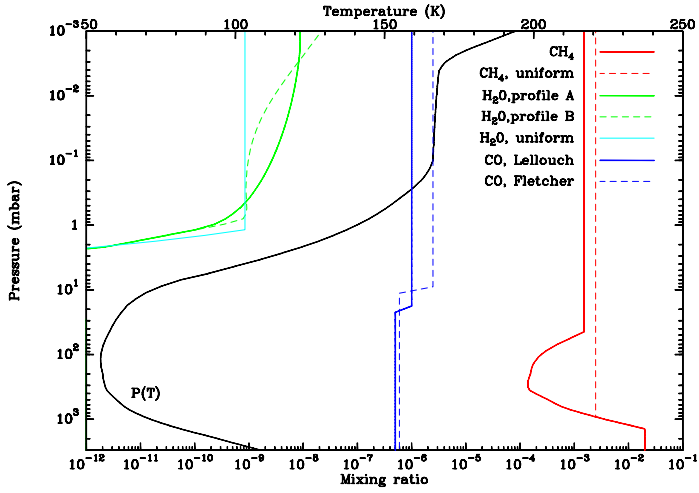}
\caption{Neptune's temperature and abundance profiles. CH$_4$ profiles condensing (thick red line) or not (thin red line) in the stratosphere 
are considered.
For H$_2$O, profiles A and B are those of Feuchtgruber et al. (1997), multiplied by 0.95 and 0.9, respectively,
and the "uniform" profile has a mixing ratio of 0.85 ppb above the condensation level.
For CO, the profiles of Lellouch et al. (2005) and Fletcher et al. (2010) are shown. The black line shows the inferred temperature
profile.}
\label{fig:thermal}
\end{figure}

\subsection{Thermal profile and D/H abundance}
Observations were analyzed by means of a standard radiative transfer code, in which the outgoing radiance from Neptune was integrated
over all emission angles. The effective spectral resolution as a function of wavelength was determined by fitting the widths of the H$_2$O
lines, whose profile is purely instrumental. We initially
considered thermal profiles inferred in previous work (Marten et al. 2005, B\'ezard et al. 1998, Fletcher et al. 2010, respectively
from ground-based, ISO, and AKARI observations). Below about 0.5 bar,
all of them follow the {\em Voyager}  radio-occultation profile (Lindal 1992, see also Moses et al. 2005). Above this level, these profiles diverge significantly, showing excursions of $\sim$5 K over 10-200 mbar, and even larger dispersion ($\sim$10-20 K) at lower pressures. Over 50-200 $\mu$m, Neptune's continuum
is formed near the 500 mbar level (T$_B$ $\sim$ 59 K). The HD lines typically probe the
10-500 mbar range (peak contribution near 2 mbar at line center). Because they show a contrasted absorption/emission appearance and because HD is vertically well mixed, 
they provide a sensitive thermometer in this region. For HD, we used the same linestrengths as in Feuchtgruber
et al. (1999). 
We found that the Fletcher et al. (2010) nominal profile (their Fig. 5) allowed a much better 
fit of the HD lines than the other two profiles, and achieved optimum fit for temperatures equal to 0.9$\times$Fletcher + 0.1$\times$Marten
(Fig. 2).
This gives 54.5 K at the tropopause, $\sim$3 K higher than in Lindal (1992). Given Neptune's temperature field as inferred from {\em Voyager} 
measurements (Conrath et al. 1998), this is probably related to the high latitude (42$^{\circ}$S) of the 
{\em Voyager} occultations. Based on mid--infrared measurements of ethane, Hammel et al. (2006) also found enhanced temperatures
(but at sub-mbar levels) compared to {\em Voyager}, a likely consequence of seasonal variability. 
Although the HD lines do not constrain temperatures above the 1 mbar level (needed in particular for analyzing the H$_2$O lines), we retained the
0.9$\times$Fletcher + 0.1$\times$Marten combination for all levels. We determined HD/H$_2$ = (9$\pm$2)$\times$10$^{-5}$, i.e. a D/H ratio of (4.5$\pm$1)$\times$10$^{-5}$ (Fig. 3). This is nominally less than but consistent with the (6.5$^{+2.5}_{-1.5})\times$10$^{-5}$ value inferred by Feuchtgruber et al. (1999) from observations of the R(2) line of HD by ISO/SWS,
and confirms that Neptune is enriched in deuterium compared to the protosolar value ($\sim$2.1$\times$10$^{-5}$) represented by Jupiter and Saturn (Lellouch
et al. 2001). We defer a joint analysis of ISO and {\em Herschel} data to future work.

\begin{figure}[ht]
\centering
\includegraphics[width=6.4cm,angle=-90]{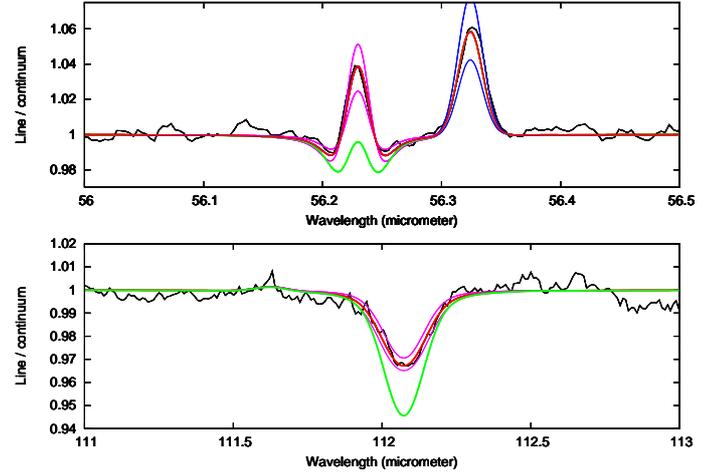}
\caption{Neptune's spectrum in the 56.0--56.5 and 111--113 $\mu$m ranges, showing HD lines at 56.25 $\mu$m (R(1)) and 112.1 $\mu$m (R(0)). Thick red line: model for HD/H$_2$ = 9$\times$10$^{-5}$ and the
nominal thermal profile of Fig. \ref{fig:thermal}. Thin pink lines: same for HD/H$_2$ = 6$\times$10$^{-5}$ and 12$\times$10$^{-5}$.
Green: model for HD/H$_2$ = 9$\times$10$^{-5}$ and Marten's et al. (2005) thermal profile).
Note also the water line at 56.35 $\mu$m, well fitted by profile A in Fig. 2. The upper and lower blue lines show models for this H$_2$O profile multiplied
and divided by 1.5.}
\label{fig:hd}
\end{figure}

\subsection{Methane, water, and carbon monoxide abundances and profiles}
\begin{figure}[ht]
\centering
\includegraphics[height=\hsize,angle=-90]{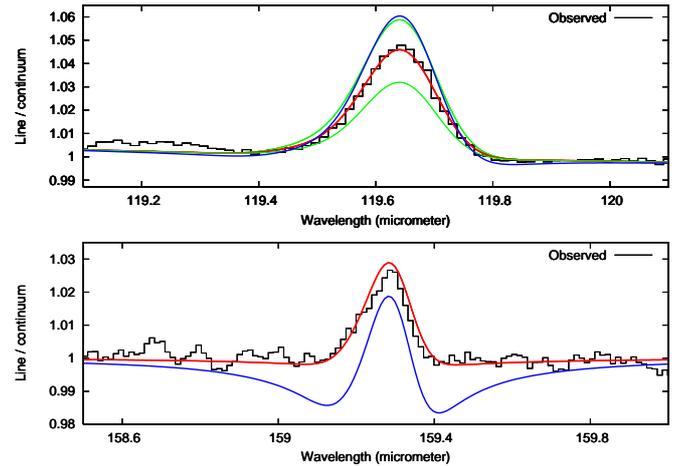}
\caption{Methane lines at 119.6 $\mu$m (Obs.ID 1342186571) and 159.3 $\mu$m (from Obs.ID 1342186537). Red: Model for stratospheric $q_{CH_4}$ = 0.0015 above the stratospheric saturation level
(thick red line in Fig. 2). Green curves: same, but for $q_{CH_4}$ = 0.0020 (upper curve) and 0.0010 (lower curve). Blue: Model in which $q_{CH_4}$ = 0.0025 down to $\sim$800 mbar (thin red line in Fig. 2).  }
\label{fig:methane}
\end{figure}

Methane has been observed in Neptune's stratosphere with a range of abundances exceeding the saturation value at the tropopause cold trap
(e.g. Baines and Hammel, 1994). The PACS spectrum shows several rotational lines of CH$_4$ in emission over 80--160 $\mu$m. 
Thanks to the mild temperature dependence of the Planck function in this spectral range, these lines are well suited to determination of the CH$_4$
stratospheric abundance. We assumed a CH$_4$ abundance of 2 \% in the deep troposphere, then following the saturation law. In the stratosphere, the CH$_4$ profile was characterized by its high--altitude mixing ratio ($q_{CH_4}$) and assumed to follow local saturation below the condensation point near 40 mbar.
Utilizing the Boudon et al. (2010) results on the absolute CH$_4$ line strengths and in 
particular using the high S/N dedicated CH$_4$ 120 $\mu$m line scan (Fig. \ref{fig:methane}),
we determined $q_{CH_4}$ = (1.5$\pm$0.2)$\times$10$^{-3}$, consistent with B\'ezard et al. (1999b) ((0.5--2)$\times$10$^{-3}$) but only marginally with Fletcher et al. (2010) ((0.9$\pm$0.3)$\times$10$^{-3}$). 
Because of the progressive increase of the continuum level longwards of 100 $\mu$m, the CH$_4$ features at 137 $\mu$m and particularly 159 $\mu$m are sensitive to the CH$_4$ amount in the lower stratosphere. An alternate assumption would be that the CH$_4$ is supersaturated there, as could perhaps result from strong convective overshoot. This situation leads, however, to unobserved absorption wings at 159 $\mu$m and to inconsistent mixing ratios for the different 
lines (Fig. \ref{fig:methane}). A 1.5$\times$10$^{-3}$ mixing ratio is $\sim$10 times greater than allowed by the 56 K cold trap, and consistent with saturation at 60 K. The most probable origin of this elevated stratospheric abundance is that CH$_4$ leaks from the hot (62--66 K at the tropopause) Southern region (Orton et al. 2007) and is redistributed planetwide by global circulation. A combined analysis of the PACS, ISO, {\em Spitzer}, and AKARI data in terms of stratospheric
methane and temperature profile will be performed in the future.

\begin{figure}[ht]
\centering
\includegraphics[width=9cm,angle=0]{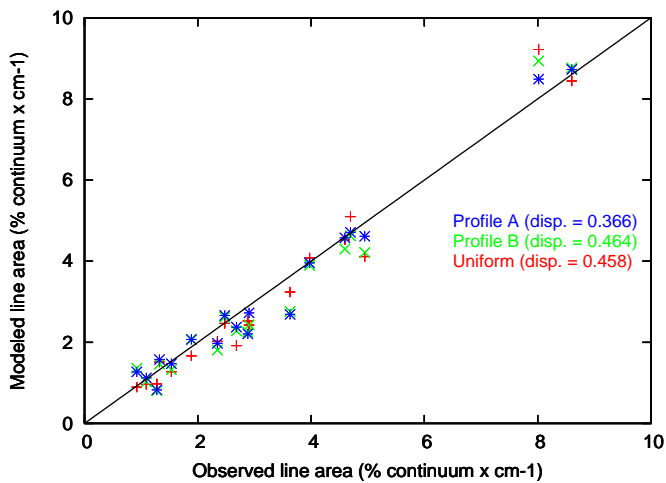}
\caption{Modeled vs. observed H$_2$O line integrated areas for the three water profiles of Fig. 2. Line areas are expressed
in cm$^{-1}$~$\times$~\% of the local continuum. For each profile, the mean rms dispersion (in the same unit) between observed and modeled 
areas is given. Profile A provides a better fit to the data than do the other two profiles.}
\label{fig:water}
\end{figure}

The presence of H$_2$O in giant planet stratospheres, including Neptune's, was established from ISO/SWS 30--45 $\mu$m spectra
(Feuchtgruber et al. 1997), demonstrating the existence of an external oxygen supply. In Neptune's case, ISO observations determined a 
(2--4)$\times$10$^{14}$ cm$^{-2}$ column density, but did not establish the water vertical profile, a parameter needed to derive the rate at which water is removed by vertical mixing and condensation and to infer the input flux of water.
More than 20 H$_2$O lines, encompassing over a range in opacity of more than an order of magnitude ($\sim$0.2 to 2.5), are detected in the PACS spectrum. If uniformly mixed
above the condensation level near 1.2 mbar, the water mixing ratio is q$_{H_2O}$ = (0.85$\pm$0.2) ppb, and its column density is 
(2.1$\pm$0.5)$\times$10$^{14}$ cm$^{-2}$. Following Feuchtgruber et al. (1997), we also considered H$_2$O vertical profiles resulting from transport models,
characterized by the eddy diffusion coefficient profile (profiles "A" and "B", see Fig. \ref{fig:thermal}). For a given vertical profile,
the water amounts we determined from the data were  identical, to within 10 \%, to the values inferred from ISO. However, the associated external fluxes
vary strongly (1.4$\times$10$^5$ cm$^{-2}$s$^{-1}$ for model A and 9$\times$10$^6$ cm$^{-2}$s$^{-1}$ for model B). We leave the detailed retrieval of
Neptune's water profile (including PACS targeted observations of several weak lines and a deep 557 GHz HIFI observation) for the future. For the time being, an elementary analysis based on the integrated linewidths favors profile A over the other two water profiles (Fig.\ref{fig:water}), suggesting
that the water mixing ratio increases with altitude over 0.1--1 mbar.

\begin{figure}[ht]
\centering
\includegraphics[width=9cm,angle=0]{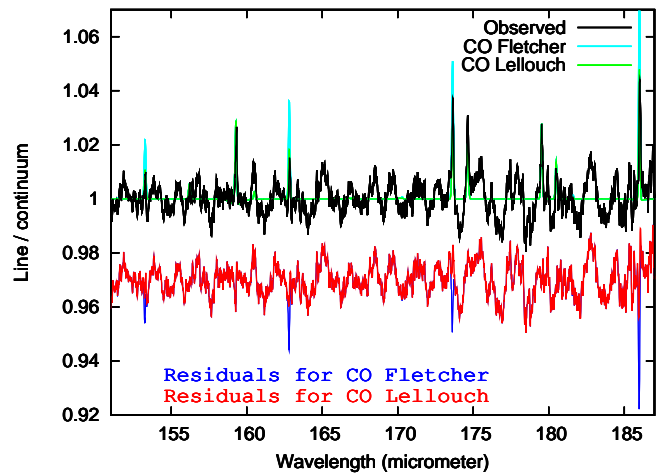}
\caption{CO lines at 153-187 $\mu$m, compared with models using the CO distributions of Lellouch et al. (2005) and Fletcher
et al. (2010), shown in Fig. \ref{fig:thermal}. CO lines occur at 154, 163, 174 and 186 $\mu$m. Other features are due
to CH$_4$ and H$_2$O. The bottom curves are difference (observed -- modeled) plots (shifted by 0.97), favoring the Lellouch et al. profile.}
\label{fig:co}
\end{figure}

Recent CO observations at millimeter/submillimeter wavelengths (Lellouch et al. 2005, Hesman et al. 2007) point to a higher abundance of CO
in Neptune's stratosphere than in the troposphere. Both studies thus indicate a dual external/internal source, with the external source possibly
provided by an ancient cometary impact. They also provide consistent values of the CO tropospheric mixing ratio (0.5-0.6 ppm). However, they differ by more than a factor of 2 (1$\times$10$^{-6}$ and 2.2$\times$10$^{-6}$, respectively) on the stratospheric CO abundance (Fig. \ref{fig:thermal}). Support for the Hesman et al. value was 
reported from the detection of CO fluorescence at 4.7 $\mu$m by AKARI (Fletcher et al. 2010), from which a 2.5 ppm abundance of CO above the 10-mbar 
pressure level was inferred. We find here that the CO lines longward of 150 $\mu$m (Fig. \ref{fig:co}) instead imply a CO stratospheric 
abundance of $\sim$1 ppm, in agreement with Lellouch et al. (2005). The detailed determination of the CO profile will be possible from combined analysis
of PACS, SPIRE, and new broadband ground-based millimeter data.
 
\begin{acknowledgements}
PACS has been developed by a consortium of institutes led by MPE
(Germany) and including UVIE (Austria); KUL, CSL, IMEC (Belgium); CEA,
OAMP (France); MPIA (Germany); IFSI, OAP/AOT, OAA/CAISMI, LENS, SISSA
(Italy); IAC (Spain). This development has been supported by the funding
agencies BMVIT (Austria), ESA-PRODEX (Belgium), CEA/CNES (France),
DLR (Germany), ASI (Italy), and CICT/MCT (Spain).
Additional funding support for some instrument activities has been
provided by ESA.
Data presented in this paper were analysed using ``HIPE'', a joint
development by the {\em Herschel} Science Ground Segment Consortium,
consisting of ESA, the NASA {\em Herschel} Science Center, and the HIFI,
PACS and SPIRE consortia. We are indebted to Bruno B\'ezard for important
discussions on the HD and CH$_4$ line parameters.
\end{acknowledgements}



\begin{thebibliography}{}
\bibitem[Baines et al. 1994]{baines94}Baines, K. \& Hammel, H. 1994, Icarus, 109, 20
\bibitem[Bezard et al. 1999b]{bezard98}B\'ezard, B., Romani, P.,   Feuchtgruber, H. \& Encrenaz, T,  1998, Ap. J., 515, 868
\bibitem[Bezard et al. 1999a]{bezard99a}B\'ezard, B., Encrenaz, T., Lellouch, E., Feuchtgruber, H., 1999a, Science, 283, 800
\bibitem[Bezard et al. 1999b]{bezard99b}B\'ezard, B., Encrenaz, T, \& Feuchtgruber, H., 1999b, ESA-SP 427, 153
\bibitem[Boudon et al. 2010]{boudon10}Boudon, V., et al., 2010, J. Quant. Spectro. Rad. Transf., 111, 1117
\bibitem[Burgdorf et al. 2003]{burgdof03}Burgdorf, M., et al., 2003, Icarus, 164, 244
\bibitem[Conrath et al. 1998]{conrath98}Conrath, B.J., Gierasch, P.J, \& Ustinov, E.A., 1998, Icarus, 135, 501
\bibitem[Feuchtgruber et al. 1997]{feucht97}Feuchtgruber, H., et al., 1997, Nature, 389, 159
\bibitem[Feuchtgruber et al. 1999]{feucht99}Feuchtgruber, H., et al., 1999, A\&A, 341, L17
\bibitem[Hammel et al. 2006]{hammel86}Hammel, H.B., et al., ApJ, 644, 1326
\bibitem[Hartogh et al. 2010]{hartogh09}Hartogh, P., et al. 2009, Planet. and Space Sci.,  57,  1596
\bibitem[Hesman et al. 2007]{hesman}Hesman, B.E., Davis, G.R., Matthews, H.E., \& Orton, G.S., 2007,
  Icarus, 186, 342
\bibitem[Lellouch et al. 2001]{lellouch01}Lellouch, E., et al., 2001, A\& A, 370, 610
\bibitem[Lellouch et al. 2005]{lellouch05}Lellouch, E., Moreno, \& G. Paubert, 2005, A\& A, 430, L37
\bibitem[Lindal 1992]{lindal92}Lindal, G.F., 1992, AJ, 103, 967
\bibitem[Fletcher et al. 2010]{fletcher10}Fletcher, L., et al. 2010, A\&A, in press
\bibitem[Marten et al. 2005]{marten05}Marten, A., et al., 2005, A\&A, 1097 
\bibitem[Meadows et al. 2008]{meadows08} Meadows, V.S., et al., 1999, Icarus, 197, 585
\bibitem[Moses et al. 2005]{moses05} Moses, J.I. et al., 2005, JGR, 110, E08001, doi:10.1029/2005JE002411
\bibitem[Orton et al. 2007]{orton07}Orton, G.S., et al., 2007, A\&A, 473, L5
\bibitem[Pilbratt et al. 2010]{pilbratt10} Pilbratt, G., \& many others 2010, A\&A this issue
\bibitem[Poglitsch et al. 2010]{poglitsch10} Poglitsch, A., Waelkens, C., Geis, N., Feuchtgruber, H., et al., 2010, A\&A this issue
\bibitem[PACS Observers Manual 2010]{pacs10} PACS Observers Manual, 2010, http://{\em Herschel}.esac.esa.int/ Docs/PACS/pdf/pacs\_om.pdf

\bibitem[]{}

%
%
%
%
%
%
%
%

\end{thebibliography}
\end{document}